\documentclass[aps, pra, 10pt, reprint]{revtex4-1}
\usepackage{amsmath}
\usepackage{amsfonts}
\usepackage{amssymb}
\usepackage{bm}
\usepackage{graphicx}

\begin{document}

\title{Quantification of memory effects in the spin-boson model}

\date{\today}

\author{Govinda Clos}
\email{govinda.clos@physik.uni-freiburg.de}

\author{Heinz-Peter Breuer}
\email{breuer@physik.uni-freiburg.de}

\affiliation{Physikalisches Institut, Universit\"at Freiburg, Hermann-Herder-Stra{\ss}e 
3, D-79104 Freiburg, Germany}

\begin{abstract}
Employing a recently proposed measure for quantum non-Markovianity, we carry 
out a systematic study of the size of memory effects in the spin-boson model
for a large region of temperature and frequency cutoff parameters. 
The dynamics of the open system is described utilizing a second-order 
time-convolutionless master equation without the Markov or rotating wave 
approximations. While the dynamics is found to be strongly non-Markovian for low 
temperatures and cutoffs, in general, we observe a special regime favoring
Markovian behavior. This effect is explained as resulting from a resonance 
between the system's transition frequency and the frequencies of the dominant 
environmental modes. We further demonstrate that the corresponding Redfield 
equation is capable of reproducing the characteristic features of the
non-Markovian quantum behavior of the model.
\end{abstract}

\maketitle

\section{Introduction}
\label{sec:intro}
Almost any realistic physical system interacts with its environment. 
The effects of this interaction can be described within the framework of open 
quantum systems \cite{TheWork,weiss}. The coupling of a quantum system to 
an environment typically leads to dissipation and decoherence processes in the
open system. However, it is known that due to the interaction, non-Markovian 
effects can emerge as well. Following the classical definition 
\cite{vankampen,NJP}, quantum non-Markovianity is often loosely described as 
the occurrence of memory effects in which the environment acts as a memory 
allowing the earlier open system states to have an effect on the later dynamics of 
the system. It is known that non-Markovian effects can play an important role in 
many applications \cite{vankampen, Kampen, gardiner}, but the absence of a 
proper definition of non-Markovianity in the quantum case impeded systematic 
studies of these effects. Recently, several formal definitions for quantum 
non-Markovianity \cite{BLP,RHP,Wolf}, which are based on 
different mathematical and physical concepts, have been proposed.

Here, we employ the definition developed in Ref.\ \cite{BLP}, which is
based on the exchange of information between the system and its environment
and enables a unique quantification of the size of quantum memory effects in the 
dynamics of a quantum system. We will study the behavior of this measure for a 
selected example: the spin-boson model, which is the paradigm of a dissipative 
two-level system \cite{leggett, weiss}.
The spin-boson model has been studied extensively over the past decades and 
has been successfully used to describe, e.g., chemical reactions, especially 
charge transfer processes in the condensed phase 
\cite{garg}, tunneling of defects and impurities in solids and 
metals \cite{golding}, and tunneling in amorphous materials 
\cite{stockburger,wuerger}. Also, the famous 
Kondo problem \cite{kondo64} could be linked to the spin-boson model 
\cite{chakravarty}, which led to the explanation for the Kondo effect 
\cite{kondo84, kondo88}. Further, the dynamics of magnetic flux 
trapped in a SQUID ring has also been understood with the help of this model and, 
based on this setup, quantum computing devices have been proposed and 
partially implemented \cite{nakamura1,nakamura2,makhlin,vion}.
The pursuit of the experimental realization of 
stable and controllable qubits has become highly important, and it is thus not 
surprising that quantum dots made of semiconductors have also been investigated 
as possible qubit implementations \cite{hayashi03,petta}.
Such quantum dots and their dissipative interaction 
with the surrounding components are often described by the spin-boson model.
Another very promising approach to quantum computation lies in trapped ions 
\cite{cirac}, as these systems allow for a very high degree of control and long qubit 
lifetimes \cite{wineland, haeffner}.
Again, the spin-boson model is a key tool to a theoretical understanding of the 
important effects \cite{porras}.

It is not only interesting to characterize memory effects in this prototypical
example, but an important motivation for the present study also lies in the need for 
a better understanding and characterization of the properties of the 
non-Markovianity measure itself, especially for more complex physical systems. 
There are several publications studying this quantity in more detail 
\cite{MeasurePaper}, applying it to physical models 
\cite{Paternostro, Mazzola,Rebentrost,Maniscalco}, 
and comparing it to other measures \cite{Kossakowski, Haikka, Zeng}. 
Additionally, several experimental measurements of the non-Markovianity measure
have been carried out recently \cite{NatPhys,EPL}, demonstrating that it provides 
an experimentally accessible observable which quantifies memory effects. 
Quite recently further theoretical implications and experimental applications to 
nonlocal quantum memory effects have been developed \cite{nl_nm}.

The paper is organized as follows: In Sec.\ \ref{sec:quant} we briefly present the 
non-Markovianity measure of Ref.\ \cite{BLP}, introduce the spin-boson 
model and the corresponding second-order time-convolutionless master equation, 
and explain the numerical simulation method used to determine the measure for
quantum non-Markovianity. The numerical simulation results are discussed and 
linked to a theoretical interpretation in Sec.\ \ref{sec:discuss}, where we also revisit 
the role of the standard Markov approximation (which is not used in Sec.\ 
\ref{sec:quant}). Finally, in Sec.\ \ref{sec:conc} we summarize our results and draw 
some conclusions.

\section{Quantifying non-Markovianity}\label{sec:quant}

\subsection{Measure for the degree of memory effects}

We quantify memory effects in the dynamics of the spin-boson system, employing
a recently proposed general measure for the degree of non-Markovian behavior in 
the evolution of an open quantum system \cite{BLP}. This measure is based 
on the idea that Markovian time evolutions are characterized by a continuous loss
of information from the open system to the environment, while non-Markovian
dynamics feature a flow of information from the environment back to the open
system. An appropriate tool to measure this information flow is given by the trace
distance between two quantum states $\rho_1$ and $\rho_2$, which is defined by 
\cite{Nielsen}
\begin{equation}
 {\cal D}(\rho_1,\rho_2)= \frac{1}{2} {\mathrm{Tr}} |\rho_1 - \rho_2|,
\end{equation}
where the modulus of an operator $A$ is given by $|A|=\sqrt{A^\dagger A}$. 
The trace distance can be interpreted as the distinguishability of the states 
$\rho_1$ and $\rho_2$ \cite{Hayashi}. We suppose that the dynamics of the open 
system is described by a family of trace preserving and completely positive 
maps $\Phi(t)$. Any pair of initial states $\rho_{1,2}(0)$ then evolves into
$\rho_{1,2}(t)=\Phi(t)\rho_{1,2}(0)$ at time $t \geq 0$. If the family of maps 
$\Phi(t)$ is divisible, e.g., if it forms a semigroup with a generator in 
Lindblad form \cite{Lindblad, Gorini}, the trace distance 
\begin{equation} \label{trace-distance-evol}
 {\cal D} (t) = {\cal D} (\rho_1(t),\rho_2(t)) 
 = {\cal D} (\Phi(t)\rho_1(0),\Phi(t)\rho_2(0))
\end{equation}
between the time-dependent pair of states decreases monotonically 
\cite{MeasurePaper}. This implies that the states $\rho_1$ and $\rho_2$ 
tend to become less and less distinguishable over time, which means that 
information about the system states is lost to the environment. On the other hand, 
whenever the trace distance increases, the two states become more 
distinguishable, and hence, information must have flowed back from the 
environment to the system, which is a clear signature for the presence of memory 
effects. 

On the basis of this interpretation, the measure for the non-Markovianity of the 
dynamics of an open system is defined by
\begin{equation}\label{eq:measure}
 {\cal N}(\Phi) = \underset{\rho_{1,2}(0)}{\operatorname{max}} \int_{\sigma>0} 
 {dt}\, \sigma(t),
\end{equation}
where $\sigma(t)=\frac{d}{dt}{\cal D} (t)$ denotes the rate of change of the 
trace distance given by Eq.\ \eqref{trace-distance-evol} and the time integration is 
extended over all intervals in which $\sigma$ is positive. The measure, thus, 
quantifies the total backflow of information from the environment to the open 
system. It further involves a maximization over the initial pair of states
$\rho_{1,2}(0)$ and therefore represents a functional of the family of dynamical 
maps $\Phi(t)$ describing the physical process. 

By definition, the measure ${\cal N}(\Phi)$ is non-negative, and we 
have ${\cal N}(\Phi)=0$ if and only if the process is Markovian. A nonzero 
measure, ${\cal N}(\Phi) > 0$, implies that the process $\Phi(t)$ is nondivisible 
and cannot be described, for example, by a time-local master equation with 
positive rates \cite{MeasurePaper}. We note that the non-Markovianity 
measure ${\cal N}(\Phi)$ represents a physically measurable quantity, as has been
demonstrated in several recent experiments \cite{NatPhys,EPL}.

\subsection{Spin-boson model and master equation}\label{sec:model}

The spin-boson model describes a two-level system which is linearly coupled to an 
environment of harmonic oscillators. The Hamiltonian of the model consists of a 
system part $H_S$, an environmental part $H_E$, and an interaction part $H_I$:
\begin{eqnarray} \label{eq:fullhamiltonian}
H &=& H_S + H_E + H_I \\
&=&
\frac{\omega_0}{2}  \sigma_z +\sum_n \left( \frac{p_n^2}{2 m_n}  
+ \frac{m_n  \omega_n^2}{2}  x_n^2  \right) 
- \frac{\sigma_x}{2} \sum_n \kappa_n x_n, \nonumber
\end{eqnarray}
where $\sigma_x$ and $\sigma_z$ are Pauli matrices, $\omega_0$ denotes 
the energy splitting of the two-level system ($\hbar$ is set equal to 1), and $m_n$, $ \omega_n $, $x_n$, and 
$p_n$ represent the masses, frequencies, and position and momentum 
operators of the environmental oscillators, respectively. Finally, $\kappa_n$ 
describes the coupling of the system to the $n$-th oscillator.

We assume that the system-environment coupling is weak and employ the
second-order time-convolutionless master equation in the interaction picture to describe the
dynamics of the open system \cite{shibata77_1,shibata77_2,Shibata,tokuyamamori},
\begin{eqnarray} \label{eq:1stafo}
\frac{d}{dt} \rho(t) &=& -i[\tilde{H}_S(t),\rho(t)] \\
  && + \sum_{i,j=0,1} a_{ij}(t) 
  \left( \sigma_i \rho(t) \sigma_j^\dagger - \frac{1}{2} 
  \left\{ \sigma_j^\dagger \sigma_i, \rho(t)\right\}\right). \nonumber
\end{eqnarray} 
We note that the derivation of this master equation is carried out without using the 
Markov or the rotating wave approximation. We have introduced the
operators $\sigma_1 = \sigma_+$ and $\sigma_0 = \sigma_-$. The 
time-dependent coefficients $a_{ij}(t)$ form a Hermitian matrix and can be 
expressed in terms of the correlation functions of the environmental operator,
\begin{equation}
B(t) = \sum_n \frac{\kappa_n}{\sqrt{2 m_n \omega_n}}
\left( a_n e^{-i \omega_n t} + a_n^\dagger e^{i \omega_n t}\right),
\end{equation}
where $a_n^\dagger$ and $a_n$ are the creation and annihilation operators of the
environmental modes. Explicitly, we find
\begin{align}
a_{11}(t)&=\frac{1}{2}  \int_0^t {ds}\, \operatorname{Re}\Big[ \mathrm{Tr}_E \{ B(t) 
B(s) \rho_E\} e ^{-i \omega_0 (t-s)}\Big], \label{eq:a11}\\
a_{00}(t)&=\frac{1}{2}  \int_0^t {ds}\, \operatorname{Re}\Big[ \mathrm{Tr}_E \{ B(t) 
B(s) \rho_E\} e ^{+i \omega_0 (t-s)}\Big],\label{eq:a00}\\
a_{10}(t)&=\frac{1}{2}  \int_0^t {ds} \, \operatorname{Re}\Big[ \mathrm{Tr}_E \{ B(t) 
B(s) \rho_E\}\Big] e ^{-i \omega_0 (t-s)}.\label{eq:a10}
\end{align}
The time-dependent system Hamiltonian
$\tilde{H}_S(t)$ is a diagonal matrix with diagonal elements
\begin{align}
h_1(t)&\!=\!\frac{1}{4} \int_0^t\! {ds}\, \operatorname{Im}\left[ \mathrm{Tr}_E \{ B(t) B(s) 
\rho_E\} e ^{+i \omega_0 (t-s)}\right],\\
h_0(t)&\!=\!\frac{1}{4} \int_0^t\! {ds}\, \operatorname{Im}\left[ \mathrm{Tr}_E \{ B(t) B(s) 
\rho_E\} e ^{-i \omega_0 (t-s)}\right].
\end{align}
The environment is assumed to be in a thermal equilibrium state initially, i.e., we
have $\rho_E = Z^{-1} e^{-\beta H_E}$, where $Z$ is the partition function and 
$\beta = \frac{1}{T}$ the inverse temperature (the Boltzmann constant is set equal
to 1). 

It is convenient to write the master equation \eqref{eq:1stafo} in terms of the
Bloch vector $\vec{v}(t)=\mathrm{Tr}_S \{ \vec{\sigma} \rho(t) \}$. Transforming back to the Schr\"odinger picture one finds \cite{TheWork}
\begin{equation}
\label{eq:vecdgl}
\frac{d}{dt} \vec{v}(t)=  M(t) \vec{v}(t)+ \vec{b}(t),
\end{equation}
where
\begin{equation}
M(t) = \left(\begin{array}{ccc}0 & - \omega_0 & 0 \\ \omega_0+ a_{yx}(t) & a_{zz}(t) 
& 0 \\0 & 0 & a_{zz}(t)\end{array}\right),
\end{equation}
and $\vec{b}(t) = \big(0,0, b_z(t)\big)^T$, with real coefficients given by
\begin{align}
\!a_{yx}(t)&\!=\! - 2 \operatorname{Im}\{a_{10}(t)\} \!=\! \frac{1}{2}\int_0^t\! ds\, D_1(s) 
\sin\omega_0s,\\
\!a_{zz}(t)&\!=\! -a_{11}(t)-a_{00}(t) \!=\! -\frac{1}{2}\int_0^t\! ds\, D_1(s) 
\cos\omega_0s,\\
\!b_z(t)&\!=\! a_{11}(t)-a_{00}(t) \!=\! -\frac{1}{2}\int_0^t\! ds\, D(s) 
\sin\omega_0s.
\end{align}
Here, $D_1(s)$ and $D(s)$ denote the noise and dissipation kernel of the
model which can be expressed in terms of the spectral density 
$J(\omega,\Omega)$ of the environmental modes as
\begin{eqnarray} 
 D_1(s) &=& 2 \int_0^\infty {d}{\omega}\, J(\omega,\Omega) 
 \coth{\left(\frac{\omega}{2 T}\right)} \cos{\omega s}, \label{eq:noisekernel} \\
 D(s) &=& 2 \int_0^\infty {d}{\omega}\, J(\omega,\Omega) \sin{\omega s}. 
 \label{eq:disskernel}
\end{eqnarray}
In the following we choose an Ohmic spectral density with Lorentz-Drude cutoff,
\begin{equation} \label{spec-density}
J(\omega,\Omega) = \frac{\gamma}{\pi} \frac{\omega}{\omega_0} 
\frac{\Omega^2}{\Omega^2 + \omega^2},
\end{equation}
where $\Omega$ is the cutoff frequency and $\gamma$ defines the coupling 
strength.

\subsection{Numerical simulation}\label{sec:num}

\begin{figure}
 \begin{center}
\includegraphics{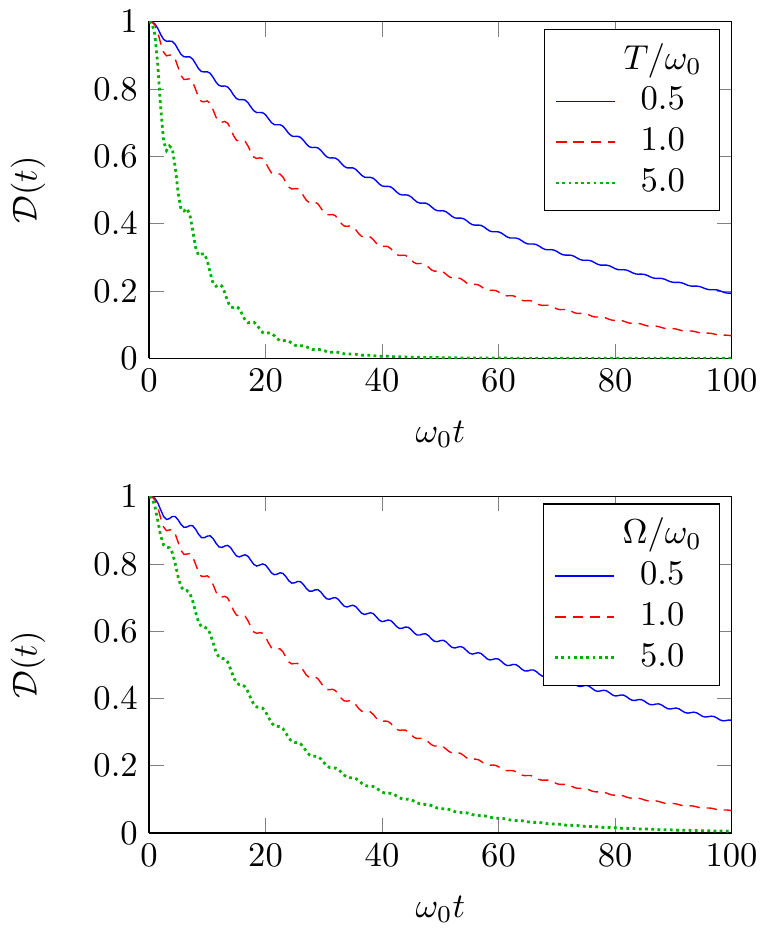}
\caption{(Color online) The trace distance as a function of time for various 
temperatures (top) and cutoffs (bottom). The parameter values of the middle curve 
in both plots are $T = \omega_0$, $\Omega=\omega_0$, 
$\gamma= 0.1 \omega_0$, and $\vec{v}_{1,2}(0) = \pm(1,0,0)^T$. The other 
curves show the effect of a variation of \emph{one} of these parameters.}
\label{fig:trdis}
 \end{center}
\end{figure}

We have carried out numerical simulations to determine the non-Markovianity
measure \eqref{eq:measure} for the spin-boson model. For any given pair of initial
states $\rho_{1,2}(0)$, represented by the corresponding Bloch vectors 
$\vec{v}_{1,2}(0)$, one first solves the Bloch equations \eqref{eq:vecdgl} 
to determine the behavior of the trace distance by means of
\begin{equation} \label{trace-distance-evol-bloch}
 {\cal D} (t) = {\cal D} (\rho_1(t),\rho_2(t)) 
 = \frac{1}{2} \left| \vec{v}_1(t) - \vec{v}_2(t) \right|,
\end{equation}
where $\left| \vec{v}_1 - \vec{v}_2 \right|$ denotes the Euclidean distance.
An example is depicted in Fig.\ \ref{fig:trdis}.

To obtain the total backflow of information given by the time integral in 
Eq.\ \eqref{eq:measure}, we subdivide
the time axis into sufficiently small intervals $[ t_{i}, t_{i+1}]$ and evaluate
the quantity
\begin{equation*} 
{\cal N}' = \frac{1}{2}\sum_i \Big[ \left| \vec{v}_1(t_{i+1})-\vec{v}_2(t_{i+1}) \right| 
- \left| \vec{v}_1(t_i)-\vec{v}_2(t_i) \right| \Big],
\end{equation*}
where the sum is extended over those intervals in which the trace distance 
increases.
The measure \eqref{eq:measure} is then found by maximizing ${\cal N}'$
over all pairs of initial states. We have realized this maximization procedure 
numerically by a Monte Carlo sampling, drawing independent random initial state 
pairs from a uniform distribution over the state space.

\begin{figure}[htbp]
\begin{center}
\includegraphics{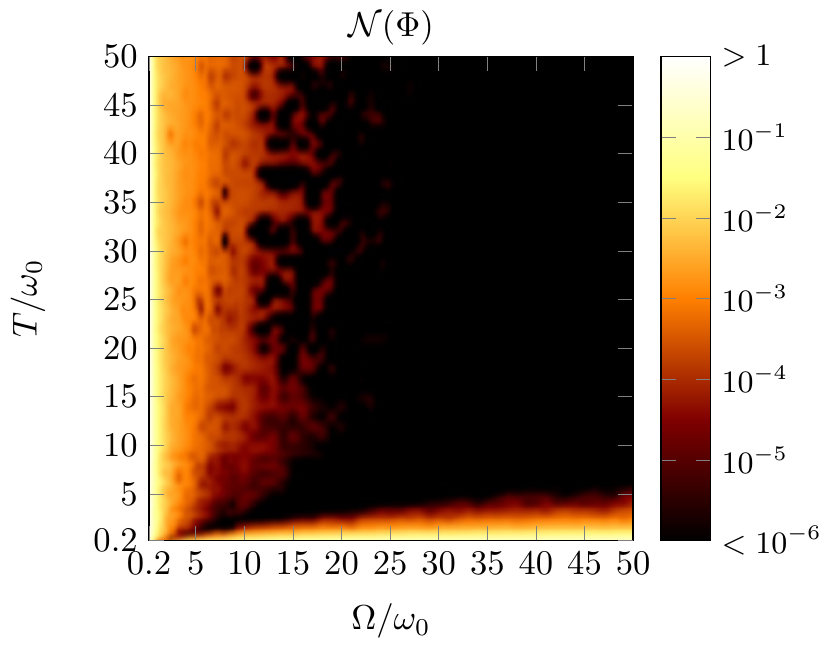}
\caption{(Color online) Non-Markovianity ${\cal N}(\Phi)$ as a function of 
$\Omega$ and $T$ up to $\Omega, T = 50 \omega_0$, indicated by the 
logarithmic colorbar.}
\label{fig:NM}
\end{center}
\end{figure}

\begin{figure}[htbp]
\begin{center}
\includegraphics{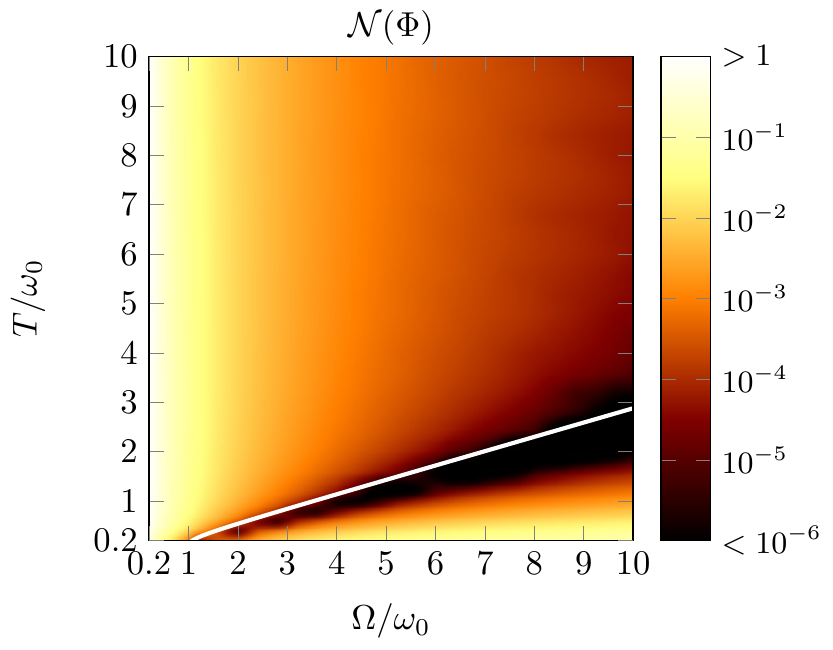}
\caption{(Color online) Non-Markovianity ${\cal N}(\Phi)$ as a function of 
$\Omega$ and $T$ up to $\Omega, T = 10 \omega_0$. For this range, the 
maximization has been refined with additional initial state pairs to yield a higher 
resolution. The white curve shows the parametrization of the resonance effect 
given by Eq.\ \eqref{eq:resonance}.}
\label{fig:NMzoom}
\end{center}
\end{figure}

Our goal is to study the behavior of the non-Markovianity as a function of 
parameters of the environment, namely, the temperature $T$ and the cutoff 
frequency $\Omega$ in the spectral density. To this end, we numerically estimate 
the measure ${\cal N}(\Phi)$ according to the above procedure for different values 
of these parameters. The result is a full characterization of the non-Markovian 
behavior in the spin-boson model depending on the properties of its environment 
depicted in Fig.\ \ref{fig:NM}.
For the preparation of this figure, the 
non-Markovianity measure has been evaluated on a grid consisting of roughly 
5000 parameter combinations, with a denser sampling in the range of 
$T,\Omega < 10 \omega_0$ shown in Fig.\ \ref{fig:NMzoom}. For each of these 
points, the Monte Carlo simulation to approximate the maximization over all state 
pairs has been done with 12000 random initial state pairs. Comparisons and tests 
with special fixed initial state pairs suggest that the speckled structure in 
Fig.\ \ref{fig:NM} is due to the maximization procedure and does not mirror a 
physical effect. We will now discuss and interpret these results in detail.
\section{Discussion and theoretical interpretation}\label{sec:discuss}

\subsection{Resonance effect}\label{sec:res}

Figure \ref{fig:NM} shows the non-Markovianity defined by Eq.\ \eqref{eq:measure} 
as a function of temperature $T$ and environmental cutoff frequency 
$\Omega$. Note that the values of the measure ${\cal N}(\Phi)$ are represented
by a logarithmic colorbar for better visualization. Based on this plot, one 
can decide for which parameters the system behaves Markovian and for which it 
exhibits strong non-Markovian effects. We observe that, roughly speaking, 
for small cutoffs $\Omega \lesssim  \omega_0$ and for small temperatures 
$T \lesssim \omega_0$ the dynamics is strongly non-Markovian, as can be seen 
from the high values of the measure ${\cal N}(\Phi)$ in these regions.
By contrast, for large values of both $\Omega$ and $T$, the measure is 
zero and the dynamics is Markovian, in full agreement with the well-known 
quantum optical limit  \cite{TheWork}. 

Our simulation results show that the non-Markovianity measure ${\cal N}(\Phi)$ 
features a characteristic structure: It exhibits a distinct minimum, i.e., a Markovian 
area within a non-Markovian regime for certain values of temperature and cutoff. 
This can be seen more clearly in Fig.\ \ref{fig:NMzoom} which shows a 
magnification of the range $\Omega, T < 10 \omega_0$ obtained from a numerical 
simulation with higher resolution. In the following we develop a physical 
interpretation for this feature which obviously depends on the temperature and the 
cutoff frequency so that it cannot be explained by looking at the spectral density $J(\omega, \Omega)$ 
only. The position of the minimum can be extracted from the Monte Carlo
simulation data and the dependence is found to be nonlinear, contrary to first the 
impression.

\begin{figure}[htb]
\begin{center}
\includegraphics{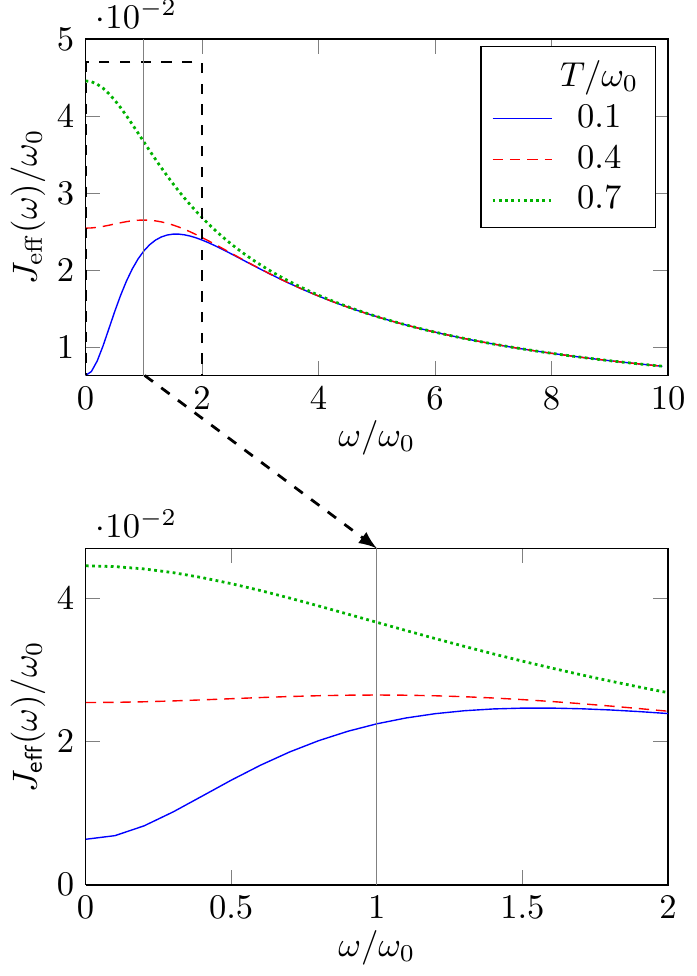}
\caption{(Color online) The effective spectral density $J_{\textsf{eff}}(\omega)$ for a fixed cutoff 
$\Omega=1.55\omega_0$ and three different values of temperature $T$. The 
middle curve ($T=0.4\omega_0$) shows the spectral density obtained from the 
resonance condition \eqref{eq:resonance}; the other two curves correspond to 
detuned temperature values. The vertical line indicates the system's transition 
frequency $\omega_0$. The bottom plot shows a magnification of the top plot.}
\label{fig:jeff}
\end{center}
\end{figure}

The environmental modes enter the master equation via the correlation functions 
in Eqs.\ \eqref{eq:a11}, \eqref{eq:a00}, and \eqref{eq:a10}. In particular, the 
coefficients $a_{yx}(t)$  and $a_{zz}(t)$ in the master equation depend on the 
noise kernel  $D_1(s)$ given in Eq.\ \eqref{eq:noisekernel}, which is determined
by the effective spectral density
\begin{equation}
 J_{\textsf{eff}}(\omega,\Omega, T) = J(\omega,\Omega) \coth{\left( \frac{\omega}{2 T}
 \right)},
\end{equation}
where $J(\omega,\Omega)$ is defined by Eq.\ \eqref{spec-density}. The spectral
distribution $J_{\textsf{eff}}$ depends on three parameters: the environmental 
mode frequency $\omega$, the cutoff frequency $\Omega$, and the temperature 
$T$. One can easily see that $J_{\textsf{eff}}$, regarded as a function of $\omega$, has 
exactly one maximum at some frequency $\omega_{\rm max}$ and the dominant 
environmental modes are located around this frequency value. Under the
resonance condition $\omega_0 = \omega_{\rm max}$ the open system sees 
an environmental power spectrum, which is approximately constant around its 
transition frequency $\omega_0$, as is illustrated in Fig.\ \ref{fig:jeff}.
In the neighborhood of the transition frequency of the system the spectral 
distribution is, thus, similar to the spectrum of white noise, and we, therefore,
expect to find the region of predominantly Markovian behavior from the condition
\begin{equation} \label{COND-1}
 \left. \frac{\partial}{\partial\omega} J_{\textsf{eff}}(\omega,\Omega,T) 
 \right|_{\omega=\omega_0} = 0.
\end{equation}
This equation defines a curve $\Omega = \Omega_{\textsf{res}}(T)$ in the
$(\Omega,T)$ plane representing the points at which the 
system's transition frequency $\omega_0$ is exactly in resonance with the 
maximum of the effective spectral density. Hence, the region of Markovian 
behavior should be located around this curve. To test this prediction we use 
Eq.\ \eqref{COND-1} to obtain the following resonance condition:
\begin{equation} \label{eq:resonance}
 \Omega_{\textsf{res}} (T) =
 \omega_0 \sqrt{\frac{T \sinh{\left( \frac{\omega_0}{T} \right)} + 
 \omega_0}{T \sinh{\left(\frac{\omega_0}{T} \right)}- \omega_0}}.
\end{equation}
This function is shown as a white curve in Fig.\ \ref{fig:NMzoom}. By 
comparison with the dark region in the figure, representing the region of very low
or even zero non-Markovianity measure ${\cal N}(\Phi)$, we see that the 
resonance condition indeed explains very well the emergence of Markovian 
behavior embedded in a region of large non-Markovianity. We note that
this picture corresponds to the results of Ref.\ \cite{BLP}, where
the damped Jaynes-Cummings model, describing the interaction of a two-level 
system with a damped cavity mode, has been studied. For this model
the non-Markovianity measure, considered as a function of the detuning 
between the system transition frequency and the frequency of the cavity mode,
shows an analogous behavior, expressing Markovian behavior for sufficiently
small detuning and a transition to non-Markovian dynamics above a certain 
threshold.

\subsection{Stationary rate approximation}\label{sec:markov}

The master equation \eqref{eq:vecdgl} can be derived by employing only the 
Born approximation, which presupposes a weak system-environment coupling.
The conventional Markov approximation consists of neglecting the 
time-dependence of the coefficients of the
master equation $a_{yx}(t)$, $a_{zz}(t)$, and $b_z(t)$ and taking their asymptotic values for $t \to \infty$ (stationary rate 
approximation). The resulting equation is also known as the Redfield master equation 
\cite{redfield57}. It provides an equation of motion with a time-independent 
generator which is not in Lindblad form and, hence, does not yield a completely 
positive quantum dynamical semigroup. However, for many physical problems the 
Redfield equation has been shown to provide a useful and accurate approximation 
of the reduced open system dynamics.
\begin{figure}[htbp]
\begin{center}
\includegraphics{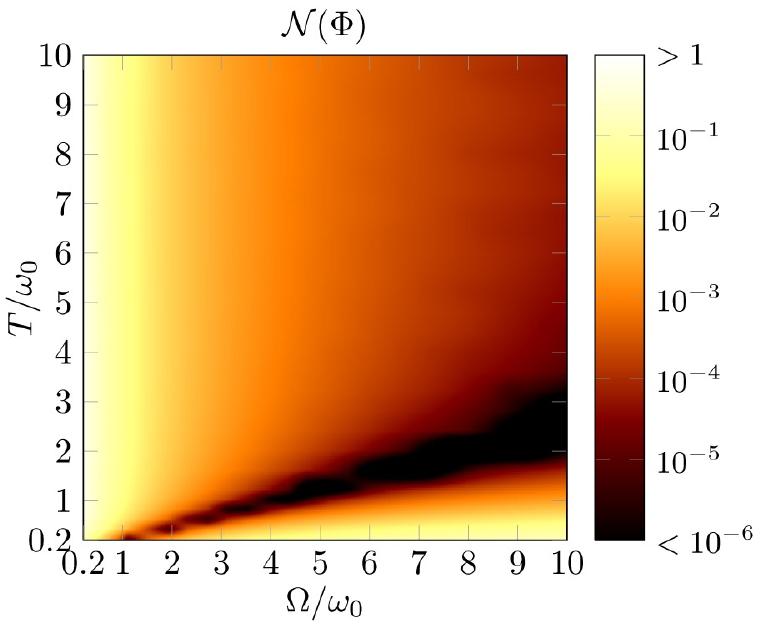}
\caption{(Color online) Non-Markovianity ${\cal N}(\Phi)$ as a function of 
$\Omega$ and $T$ up to $\Omega, T = 10 \omega_0$ for the master equation 
with stationary rates (Redfield equation).}
\label{fig:NMstat}
\end{center}
\end{figure}

We have carried out numerical simulations to determine the non-Markovianity
measure ${\cal N}(\Phi)$ under the stationary rate approximation leading to the
Redfield equation. Results are shown in Fig.\ \ref{fig:NMstat}. Quite remarkably, 
the dynamics generated by the Redfield equation is still strongly non-Markovian
in a large parameter region. Actually, comparing Figs.\ \ref{fig:NMstat} and 
\ref{fig:NMzoom} we observe that the overall behavior of the non-Markovianity 
measure is qualitatively and even quantitatively very similar to the case without 
stationary rate approximation. Thus, we conclude that the stationary rate
approximation, which is conventionally regarded as the Markov approximation, does 
not lead to a quantum Markov process in the sense of the definition for
non-Markovianity used here.

\section{Conclusions}\label{sec:conc}

In this article we have quantified non-Markovian effects in the dynamics of the 
spin-boson model depending on properties of the environment.
We have characterized the non-Markovianity of the dynamical map as a function 
of the temperature and the cutoff frequency in the spectral density. This result 
serves to classify different regimes in the parameter range. On the one  hand, we 
can understand why the predictions of a Lindblad master equation fail in some 
applications, and, on the other hand, our findings may be used as a tool to devise 
experimental setups that feature especially large non-Markovian effects.

During the maximization over all initial state pairs for the evaluation of the 
non-Markovianity measure, we found that the main contribution to the measure 
stems from the time evolution of the off-diagonal elements of the system density 
matrix, i.e., the coherences. This fact suggests that at least in our case, the 
non-Markovianity is mainly a quantum effect. Consequently, a description by 
means of a classical rate equation might fail to reproduce important memory 
effects.

It is often believed that non-Markovianity is to a large extent only a transient 
phenomenon. However, our comparison of the results for the non-Markovianity 
measure applied to the full master equation with those for the master equation with 
stationary rate approximation (Redfield master equation) reveals that this intuition 
might be misleading: Although the initial oscillations of the coefficients in the 
master equation decay over times which are short compared to the system's 
relaxation time, the non-Markovianity measure steadily builds up over the whole 
relaxation process, showing no special significance of the transient time evolution. 
Consequently, this approximation yields qualitatively similar results for the 
measure, despite the fact that it neglects any transient effects.

In this context, the meaning of the standard Markov approximation must be 
revisited. The approximation of stationary rates in the master equation is often
regarded as Markovian description and the rates are called Markovian rates. 
However, our results demonstrate that an approximation of this kind does not 
necessarily lead to a Markovian dynamics in the sense of our definition in terms of
the information flow. In fact, we found that the master equation with
stationary rates is still capable of describing the essential features of the 
non-Markovian behavior. It should also be noticed in this context that in the
derivation of a Lindblad master equation the rotating wave approximation is a 
crucial step in order to obtain a completely positive quantum dynamical semigroup.
Thus, it is the combination of rotating wave and standard Markov approximation 
that has the desired effect of yielding a Markovian semigroup; the term 
{\textit{Markov approximation}} can in this sense be deceptive.

The dynamics given by the master equation \eqref{eq:vecdgl} without the rotating 
wave and stationary rate approximations not only shows strong non-Markovian 
behavior for small temperatures and/or environmental cutoff frequencies, but also 
features an interesting resonance effect: For a certain narrow range of parameters 
within the non-Markovian regime, the system exhibits Markovian dynamics. We 
were able to explain this effect by showing that this Markovian behavior occurs 
when the system transition frequency is in (or close to) resonance with the 
dominant modes of the environment. This can be understood by recognizing that 
in this case the distribution of modes in the vicinity of the system frequency 
resembles a uniform distribution (white noise), which clearly favors Markovian 
behavior. The next step towards a deeper understanding of the resonance effect is 
to study the behavior of the non-Markovianity measure for other models involving, 
in particular, structured reservoir spectral densities with several maxima or spectral 
gaps.

\acknowledgments

We thank Jyrki Piilo and Elsi-Mari Laine for interesting discussions and
helpful comments on the manuscript. Financial support by the German Academic 
Exchange Service (DAAD) is gratefully acknowledged.

\end{document}